\documentclass[superscriptaddress, reprint, amsmath, amssymb, aps, prx, 10pt, twocolumn, floats]{revtex4-2}

\usepackage[english]{babel}
\usepackage[utf8]{inputenc}
\usepackage[T1]{fontenc}

\usepackage{graphicx}
\usepackage[caption=false]{subfig}
\usepackage{float}
\usepackage{dcolumn}
\usepackage{blindtext}
\usepackage{upgreek}
\usepackage{siunitx}
\usepackage{xcolor}

\DeclareSIUnit\gauss{G}
\usepackage{braket} 

\newcommand{\kcirc}[1]{\ket{#1 \mathrm{C}}}
\newcommand{\hsigma}{{\hat{\sigma}}}

\newcommand{\Som}{0_{m,s}}
\newcommand{\Sos}{0_{m,s}}
\newcommand{\Sim}{1_m}
\newcommand{\Sis}{1_s}
\newcommand{\Soa}{0_a}
\newcommand{\Sia}{1_a}

\newcommand{\kom}{\ket{\Som}}
\newcommand{\kos}{\ket{\Sos}}
\newcommand{\kim}{\ket{\Sim}}
\newcommand{\kis}{\ket{\Sis}}
\newcommand{\koa}{\ket{\Soa}}
\newcommand{\kia}{\ket{\Sia}}

\newcommand{\vv}[1]{\boldsymbol{#1}}
\newcommand{\e}{\mathrm{e}}

\newcommand{\hU}{\hat{\mathcal{U}}}

\usepackage{wasysym}
\usepackage{enumitem}
\usepackage{bm}

\begin{document}

\title{Spin-exchange interactions between circular Rydberg atoms over long times}

\author{A.~Durán-Hernández}
\altaffiliation{These authors contributed equally to this work}
\author{G.~Creutzer}
\altaffiliation{These authors contributed equally to this work}
\author{A.~A.~Young}
\author{A.~Kassid}
\author{Y.~Machu}
\altaffiliation[Current address: ]{Pasqal, 24 rue Emile Baudot - 91120 Palaiseau, Paris, France}
\author{J.~M.~Raimond}
\author{M.~Brune} 
\affiliation{Laboratoire Kastler Brossel, Coll\`ege de France, CNRS, ENS-Universit\'e PSL, Sorbonne Universit\'e, 11 place Marcelin Berthelot, F-75231 Paris, France}
\author{C.~Sayrin}
\email[Corresponding author: ]{clement.sayrin@lkb.ens.fr}
\affiliation{Laboratoire Kastler Brossel, Coll\`ege de France, CNRS, ENS-Universit\'e PSL, Sorbonne Universit\'e, 11 place Marcelin Berthelot, F-75231 Paris, France}
\affiliation{Institut Universitaire de France, 1 rue Descartes, 75231 Paris Cedex 05, France}
\date{\today}

    \begin{abstract}
\bfseries 
Arrays of neutral atoms~\cite{Barredo2016, Endres2016} excited to Rydberg levels have emerged as one of the most promising platforms for quantum computation and simulation~\cite{Browaeys2020, Morgado2021, Bluvstein2026}. With the hope to outperform classical devices, the number of atoms has been increased by orders of magnitude~\cite{Pichard2024, Manetsch2025, Lin2025, Lim2026, Zhu2026}. However, the interaction time, i.e., the maximum accumulated time during which an atom interacts with its neighbours, has been limited to a few microseconds only~\cite{Evered2023, Emperauger2025a}. This restrains the number of gates per atom or prevents the simulation of long-time dynamics of quantum many-body systems. 
Here, we observe the spin-exchange interaction between two spin 1/2s encoded in laser-trapped circular Rydberg atoms~\cite{Nguyen2018, Ravon2023, Holzl2024} over more than $\boldsymbol{60\,\upmu\mathrm{s}}$ and $\boldsymbol{40}$ spin-oscillation periods, improving the state of the art~\cite{Emperauger2025a} by an order of magnitude. These unprecedented timescales allow us to record for weak atomic trapping a collapse and revival of the spin-exchange oscillation contrast induced by spin-motion coupling~\cite{Mehaignerie2023}. Our results constitute the first observation of this coupling with laser-trapped Rydberg atoms. To counteract its detrimental effect for quantum simulation, we demonstrate a novel dynamical decoupling method that prevents the collapse of the spin oscillations. This method exploits the advantages of our recently developed hybrid platform that enables the measurement and the optical manipulation of long-lived circular Rydberg atoms with auxiliary Rydberg atoms~\cite{Machu2026}. This work opens a direct route to long-duration quantum simulation of strongly-interacting many-body systems. 
    \end{abstract}

\maketitle 
By emulating condensed matter systems on fully-controllable experimental platforms, quantum simulators promise to access rich phenomena that are hard to grasp through classical simulations~\cite{Feynman1982, Lloyd1996}. Powerful numerical methods~\cite{Wu2024a} already enable deep understanding of ground-state properties of interacting spin systems and many of these predictions have been confirmed using arrays of Rydberg atoms~\cite{Browaeys2020, Bai2026a}. They have led to the simulation of complex phases of matter and of quantum phase transitions, with seminal observations of antiferromagnetic ordering~\cite{Scholl2021, Ebadi2021} and of topological phases~\cite{deLeseleuc2019, Semeghini2021, Evered2025, Bornet2026}. Rydberg-based quantum simulators have also been used to observe early-time quantum dynamics, driven by Ising or XY Hamiltonians following quantum phase transitions~\cite{Semeghini2021, Chen2025a, Manovitz2025}, revealing the onset of scarring-like~\cite{Bluvstein2021} and string-breaking dynamics~\cite{Gonzalez-Cuadra2025} as well as of quantum information scrambling~\cite{Liang2025}. 

Quantum simulations are of particular interest in the regimes that cannot be simulated classically. Those can be reached, on the one hand, by increasing the number $N$ of interacting Rydberg atoms. Most recent experiments feature a few thousand atoms laser trapped in their ground states in optical tweezers~\cite{Pichard2024, Manetsch2025, Lin2025, Lim2026, Zhu2026} and have led to quantum simulations with up to $N\gtrsim200$ atoms~\cite{Scholl2021, Ebadi2021, Semeghini2021, Leclerc2026}. On the other hand, quantum simulators would be highly valuable when operated over extended periods of time, even with moderate numbers of atoms~\cite{Schmitt2022, Haghshenas2026, Vovrosh2026, Tindall2026}. They would enable the study of complex phenomena including ergodicity breaking through quantum scarring~\cite{Turner2018a, Serbyn2021} or quantum chaos and scrambling of quantum information~\cite{Hosur2016,Yuan2022, Xu2024}.

Unfortunately, the duration of Rydberg-based simulations has been limited to a few microseconds, corresponding to a few characteristic interaction times only, for two main reasons. First, the relaxation time of the simulator is limited to $\tau/N$, where $\tau$ is the lifetime of the employed Rydberg levels, optically excited from the ground state, which lies in the few $\SI{100}{\micro\second}$ range. Second, the free motion of the atoms, usually untrapped during the simulation, introduces an emergent disorder, which strongly affects the many-body dynamics after a few microseconds~\cite{Dag2025}. Though laser trapping of Rydberg atoms has been recently demonstrated~\cite{Cortinas2020a, Barredo2020, Wilson2022, Ravon2023, Holzl2024}, quantum simulation with laser-trapped Rydberg atoms has, to our knowledge, never been performed. 

In this work, we observe spin-exchange dynamics emulated by two laser-trapped Rydberg atoms over $\SI{67}{\micro\second}$ and $40$ oscillation periods, improving the state of the art~\cite{Emperauger2025a} by an order of magnitude. We employ a recently developed hybrid platform where spin-1/2 particles, or qubits, are encoded into circular Rydberg levels, i.e., high-principal-quantum-number ($n$) levels with maximal angular momentum ($\ell=n-1$), and where auxiliary atoms, transiently excited to laser-accessible low-$\ell$ Rydberg levels, enable local manipulation and quantum non-demolition (QND) measurement of the qubits~\cite{Machu2026}. The duration of the simulation is in part limited by the lifetime of the circular Rydberg levels, in the $\SI{100}{\micro\second}$ range in our room-temperature setup, but which we effectively double through a post-selection technique, akin to erasure measurements~\cite{Ma2023, Scholl2023}. 

This unprecedented duration allows us to observe the influence of the coupling between spin and motional degrees of freedom. 
While this coupling has recently been studied with free-flying interacting Rydberg atoms~\cite{Bharti2024, Emperauger2025}, our results constitute its first observation for laser-trapped Rydberg atoms. In particular, with a weak enough atomic trapping, we record the collapse of spin oscillations  and their revival around a trap oscillation period. 
Finally, we demonstrate a novel spin-motion dynamical decoupling technique, which counteracts, for trapped atoms only, the motion-induced collapse of the oscillations. 

\begin{figure}
\centering
\includegraphics[width=0.99\linewidth]{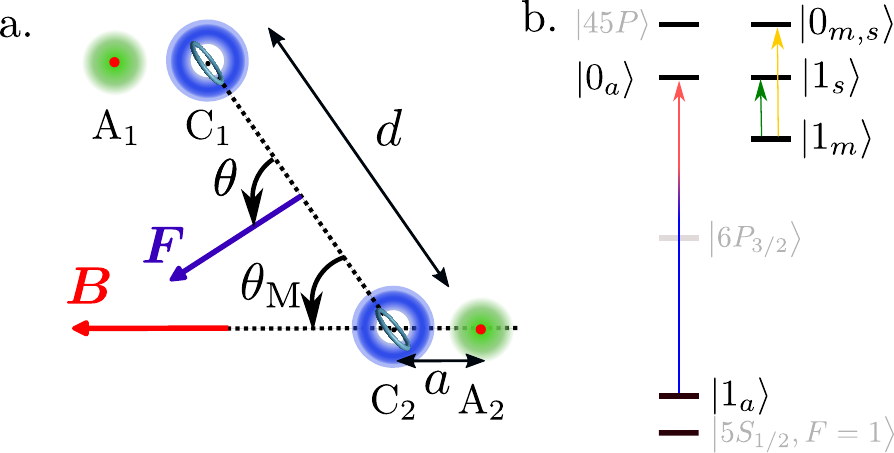}
	\caption{\textbf{Manipulation and measurement of interacting circular Rydberg qubits.} 
	\textbf{a.} We trap two circular Rydberg atoms ($\mathrm{C}_1$ and $\mathrm{C}_2$, cyan tori) inside optical bottle beams (blue rings) and two auxiliary atoms ($\mathrm{A}_1$ and $\mathrm{A}_2$) in their ground state (red dots) inside Gaussian optical tweezers (green disks). The distance between the two circular Rydberg atoms is set to $d=\SI{19}{\micro\meter}$ and that between $\mathrm{A}_k$ and $\mathrm{C}_k (k\in\{1,2\})$ to $a=\SI{5.2}{\micro\meter}$. We apply a $\SI{14}{\gauss}$ magnetic field $\vv{B}$ (red arrow) parallel to the $(\mathrm{A}_k\mathrm{C}_k)$ axes, which make an angle $\theta_\mathrm{M}\approx\SI{54.7}{\degree}$ with the $(\mathrm{C}_1\mathrm{C}_2)$ axis. The electric field $\vv{F}$ (blue arrow) makes an angle $\theta$ with the latter. During the state preparation and measurement, $\theta=\theta_\mathrm{M}$ to cancel the interaction between $\mathrm{C}_1$ and $\mathrm{C}_2$ and maximize that between $\mathrm{A}_k$ and $\mathrm{C}_k$. We switch to $\theta=\pi/2$ (as sketched) during the simulation of the spin exchange to turn on the interactions between the circular Rydberg atoms. 
	\textbf{b. Simplified level structure} of Rubidium-87 atoms, displaying the relevant levels for auxiliary (left) and circular Rydberg (right) atoms. The red-and-blue arrow represents the optical excitation from the ground state $\kia=\ket{5S_{1/2}, F=2, m_F=2}$ to the low-$\ell$ Rydberg state $\koa=\ket{45S_{1/2}, m_J=1/2}$. The green and yellow arrows indicate microwave excitation to $\ket{0_{m,s}}=\kcirc{54}$ from $\kim=\kcirc{52}$ and $\kis=\kcirc{53}$, respectively. When $F=\SI{1.59}{\volt\per\centi\meter}$, a Förster resonance condition is met, for which $\ket{\Soa, \Sos}$ and $\ket{45P, \Sis}$ are degenerate. }
	\label{fig:setup}
\end{figure}

\begin{figure*}
\centering
\includegraphics[width=0.99\linewidth]{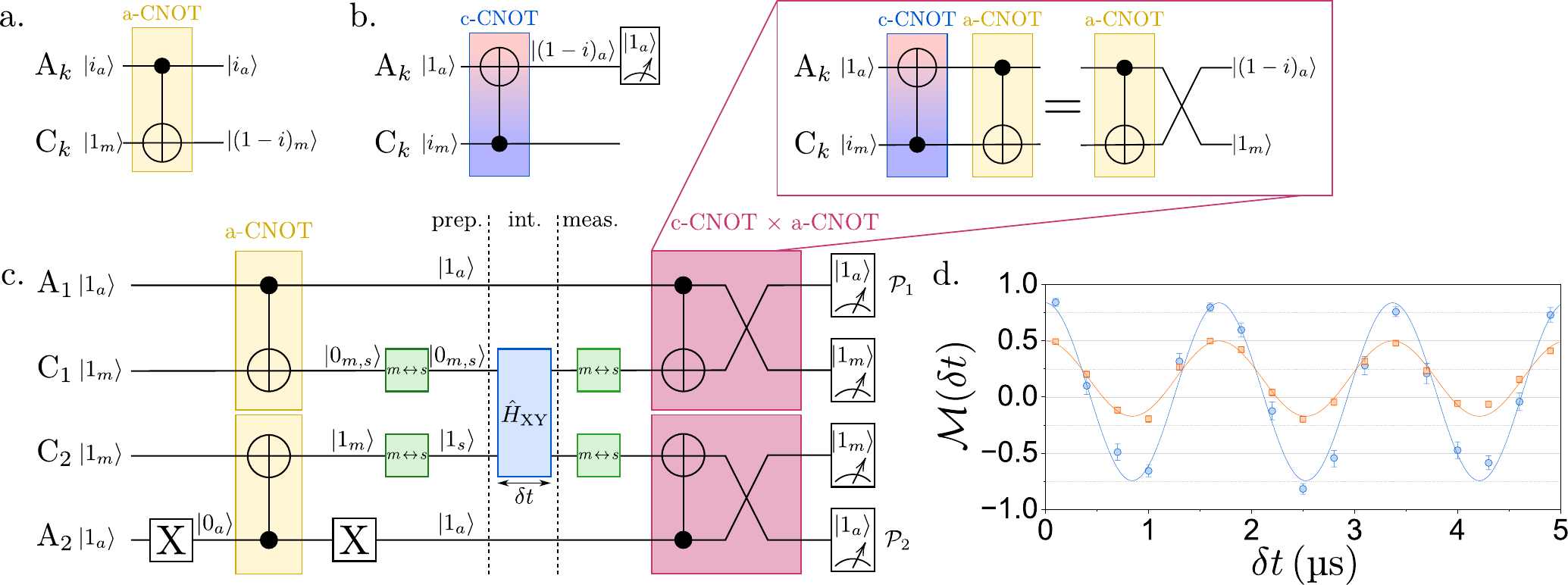}
	\caption{\textbf{Manipulation and measurement of interacting circular Rydberg qubits.} 
	\textbf{a. Quantum circuit of the circular-qubit state control.} A mw $\pi$-pulse on the $\kim\to\kom$ transition applied on a circular Rydberg qubit $\mathrm{C}_k$ in the vicinity of an auxiliary atom $\mathrm{A}_k$ in $\ket{i_a}, i\in\{0,1\}$ is equivalent to an auxiliary-atom-controlled CNOT gate (a-CNOT, yellow box). $\mathrm{C}_k$ initially in $\kim$ ends up in $\ket{(1-i)_m}$. 
	\textbf{b. Quantum circuit of the auxiliary-atom-based measurement.} The optical excitation of $\mathrm{A}_k$ from $\kia$ to $\koa$ in the vicinity of $\mathrm{C}_k$ in $\ket{i_m}, i\in\{0,1\},$ is equivalent to a circular-qubit-controlled CNOT gate (c-CNOT, red-blue box), which leaves $\mathrm{A}_k$ in $\ket{(1-i)_a}$. A subsequent fluorescence measurement of $\kia$ (white arrowed box, the measured state is indicated) performs a measurement of the circular qubit state. 
	\textbf{c. Spin-exchange measurement sequence.} The quantum circuit schematically represents the preparation, interaction and measurement stages as a series of gates applied on the auxiliary ($\mathrm{A}_1$ and $\mathrm{A}_2$) and circular ($\mathrm{C}_1$ and $\mathrm{C}_2$) qubits. We indicate the states of the atoms at all steps. The X box represents an optical $\pi$ pulse on the $\kia\leftrightarrow\koa$ transition and the green "$m\leftrightarrow s$" boxes a mw $\pi$ pulse on the $\kim\leftrightarrow\kis$ one, used to switch between the measurement and simulation bases. The blue box corresponds to the time $\delta t$ during which $\mathrm{C}_1$ and $\mathrm{C}_2$ interact under the Hamiltonian $\hat{H}_{XY}$, with $\theta$ set to $\pi/2$. We eventually measure the probabilities $\mathcal{P}_1$ and $\mathcal{P}_2$ to detect $\mathrm{A}_1$ and $\mathrm{A}_2$, respectively, in $\kia$. Before the measurement, we perform an a-CNOT and a swap gate (pink box). Ideally, it resets the state of the circular qubits to $\kim$ but is equivalent to the auxiliary-atom-based measurement (panel b) for the auxiliary atoms. The detection of the circular qubits in $\kim$ is used for the syndrome measurement.	
	\textbf{(inset) Syndrome measurement.} Adding an a-CNOT gate after the c-CNOT gate of the auxiliary-atom-based measurement is equivalent to performing an a-CNOT gate followed by a swap gate. $\mathrm{C}_k$ initially in $\ket{i_m}$ is always left in $\kim$ while $\mathrm{A}_k$ ends up in $\ket{(1-i)_a}$, as with the c-CNOT gate only. A post-selection of the experimental realizations on the detection of $\mathrm{C}_k$ in $\kim$ is used to remove preparation and some relaxation-induced errors.	
	\textbf{d. Spin-exchange interaction.} Population imbalance $\mathcal{M}(\delta t) = \mathcal{P}_1(\delta t)-\mathcal{P}_2(\delta t)$ measured with the sequence of panel c, with (blue circles) or without (orange squares) post-selection on the measurement of the circular qubits in $\kim$. The solid lines are sinusoidal fits to the data, with a common frequency $J = \SI{594\pm2}{\kilo\hertz}$ and amplitudes $A=\num{0.79\pm0.03}$ (blue) and $\num{0.33\pm0.01}$ (orange). The errors bars are Clopper-Pearson confidence intervals.
	}
	\label{fig:circuit}
\end{figure*}

\section*{Hybrid Rydberg quantum simulator}
Our experimental platform has recently been described in Ref.~\cite{Machu2026} and is sketched in Extended Fig.~\ref{edfig:setup} (see also methods). We excite a pair of atoms ($\mathrm{C}_1$ and $\mathrm{C}_2$, see Fig.~\ref{fig:setup}.a) to circular Rydberg levels and laser trap them in optical bottle beams (BoBs)~\cite{Ravon2023}, with an interatomic distance $d$ of $\SI{19}{\micro\meter}$. We encode the qubits using three different circular Rydberg states: $\kim\equiv\kcirc{52}$ and $\ket{0_m}\equiv\kcirc{54}$ as a measurement basis, $\kis\equiv\kcirc{53}$ and $\ket{0_s}=\ket{0_m}$ (denoted $\kom$ in the following) as a simulation basis, where $\kcirc{n}$ denotes the circular Rydberg state with principal quantum number $n$ (see level structure in Fig.~\ref{fig:setup}.b). We initially prepare atoms $\mathrm{C}_1$ and $\mathrm{C}_2$ in $\kim$, using optical, radiofrequency and microwave (mw) excitations within a constant $\SI{14}{\gauss}$ magnetic field, $\vv{B}$, and a dynamically-controllable electric field, $\vv{F}$, initially parallel to $\vv{B}$~\cite{Ravon2023}. The two other circular states can be reached from $\kim$ by mw excitation. In the following, we denote $\ket{\psi, \phi}$ the pair state where atom $\mathrm{C}_1$ is in $\ket{\psi}$ and atom $\mathrm{C}_2$ in $\ket{\phi}$. Spin-exchange oscillations can only be observed when preparing the atoms in two different states, here $\kos$ and $\kis$. This requires local manipulation and detection of the circular Rydberg atoms, provided by our recently-demonstrated hybrid Rydberg platform~\cite{Machu2026}.

In the vicinity of $\mathrm{C}_1$ and $\mathrm{C}_2$, we trap auxiliary atoms $\mathrm{A}_1$ and $\mathrm{A}_2$ in their ground state $\kia\equiv\ket{5S_{1/2}, F=2, m_F=2}$ in Gaussian optical tweezers. We transiently optically excite them to a low-$\ell$ Rydberg level, $\koa\equiv\ket{45S}$. We make use of a Stark-tuned Förster resonance to engineer a strong first-order dipole-dipole interaction between an auxiliary atom in $\koa$ and a neighbouring circular atom in $\kom$~\cite{Cohen2021}, a method also employed in dual-species~\cite{Beterov2015, Anand2024, Miles2026, Wang2026} or Rydberg-molecule~\cite{Zhang2022, Zhu2025, Ruttley2026} experiments. Atoms $\mathrm{A}_k$ and $\mathrm{C}_k$, $(k\in\{1,2\})$, are aligned with $\vv{B}$ and the initial electric field to maximize the interaction energy, of $h\times\SI{32}{\mega\hertz}$, where $h$ is Planck's constant, for the set interatomic distance of $a=\SI{5.2}{\micro\meter}$. Cross interactions between auxiliary atoms $\mathrm{A}_{1,2}$ and circular Rydberg atoms $\mathrm{C}_{2,1}$ are negligible (see methods).

The auxiliary atoms can control the state of the circular Rydberg atoms. An auxiliary atom $\mathrm{A}_k$ in $\koa$ prevents the mw-induced transfer of $\mathrm{C}_k$ from $\kim$ to $\kom$~\cite{Machu2026}. This state-dependent blockade~\cite{Lukin2001a} of a $\SI{0.8}{\micro\second}$-long mw $\pi$ pulse on the $\kim\to\kom$ transition corresponds to an auxiliary-qubit-controlled not (a-CNOT) gate on the circular qubit (Fig.~\ref{fig:circuit}.a). With $\mathrm{A}_k$ in $\koa$ or $\kia$, $\mathrm{C}_k$ initially in $\kim$ ends up in $\kim$ or $\kom$, respectively. In short, the state of $\mathrm{A}_k$ ($\ket{i_a}, i\in\{0,1\}$) is effectively anti-mapped onto that of $\mathrm{C}_k$, which becomes $\ket{(1-i)_m}$. 

Similarly, the Rydberg blockade of the optical excitation ($\SI{0.3}{\micro\second}$-long $\pi$ pulse) of $\mathrm{A}_k$ from $\kia$ to $\koa$ when $\mathrm{C}_k$ is in $\kom$ enables a circular-qubit-controlled not (c-CNOT) gate on the auxiliary qubit (Fig.~\ref{fig:circuit}.b). The state of $\mathrm{C}_k$ ($\ket{i_m}, i\in\{0,1\}$) is anti-mapped onto that of $\mathrm{A}_k$, which becomes $\ket{(1-i)_a}$. A subsequent fluorescence detection of the ground state $\kia$ performs a QND measurement of the circular Rydberg atom in $\kom$~\cite{Machu2026}. The probability to detect $\mathrm{A}_k$ in $\kia$, denoted $\mathcal{P}_k$, is the probability to detect atom $\mathrm{C}_k$ in $\kom$.

\section*{State preparation and measurement}
In order to observe a spin-exchange interaction between the two circular Rydberg atoms, we need to prepare them in different qubit states. The corresponding quantum circuit is sketched in Fig.~\ref{fig:circuit}.c. First, we prepare $\mathrm{C}_1$ and $\mathrm{C}_2$ in $\kim$ and the auxiliary atoms in $\kia$. Then, we selectively excite $\mathrm{A}_2$ to $\koa$ (first X gate in Fig.~\ref{fig:circuit}.c), perform the a-CNOT gate on both atom pairs, followed by the de-excitation of $\mathrm{A}_2$ to $\kia$ (X gate). This leaves the circular-Rydberg pair in the $\ket{\Som, \Sim}$ state and both auxiliary atoms in $\kia$. 

The interaction between two atoms in $\kom$ and $\kim$ is a weak second-order van-der-Waals interaction, with a spin-exchange frequency of a few hertz when $d\sim\SI{20}{\micro\meter}$ (see methods). To get a strong interaction between $\mathrm{C}_1$ and $\mathrm{C}_2$ requires us to switch from the measurement to the simulation basis (green $m\leftrightarrow s$ box in Fig.~\ref{fig:circuit}.c), i.e., to transfer the atom in $\kim$ to $\kis$, leaving the circular atom pair in $\ket{\Sos, \Sis}$. The first-order dipole-dipole interaction between these states maps onto the XY interaction Hamiltonian, which reads
\begin{align}
	\hat{H}_\mathrm{XY} = \frac{h J(d)}{2}\, \left(3\cos^2\theta - 1\right)\, \left(\hsigma_1^+\hsigma_2^- + \hsigma_1^-\hsigma_2^+\right) \ ,
	\label{eq:HXY}
\end{align} 
where the $\hsigma_k^\pm$s are the Pauli matrices in the simulation basis for atom $\mathrm{C}_k$. The interaction strength $J(d)$ scales as $1/d^3$ and amounts to $\SI{0,56}{\mega\hertz}$ when $d=\SI{19}{\micro\meter}$~\cite{Mehaignerie2025}. The angle $\theta$ is the angle between the interatomic axis and the quantization axis, aligned with $\vv{F}$ (Fig.~\ref{fig:setup}.a). 

During the circular-state preparation and the a-CNOT and c-CNOT gates, $\theta$ is set to the \emph{magic} value $\theta_\mathrm{M}\approx \SI{54.7}{\degree}$ that cancels interactions between $\mathrm{C}_1$ and $\mathrm{C}_2$~\cite{Mehaignerie2025, Huls2026}. Once the latter are prepared in $\ket{\Sos, \Sis}$, we turn on their mutual interaction by rotating the electric field to $\theta=\SI{90}{\degree}$. The spin-exchange frequency between the atoms suddenly switches from $0$ to $J(d)$. 
We let $\mathrm{C}_1$ and $\mathrm{C}_2$ interact over a duration $\delta t$ (blue box in Fig.~\ref{fig:circuit}.c), before rotating the electric field back to the interaction-free ($\theta=\theta_\mathrm{M}$) configuration. Then, we switch back to the measurement basis by transferring atoms in $\kis$ to $\kim$ (green $m\leftrightarrow s$ box in Fig.~\ref{fig:circuit}.c) and we measure the state of the circular Rydberg atoms using the auxiliary atoms.

\begin{figure*}[t]
\centering
	\includegraphics[width=0.95\linewidth]{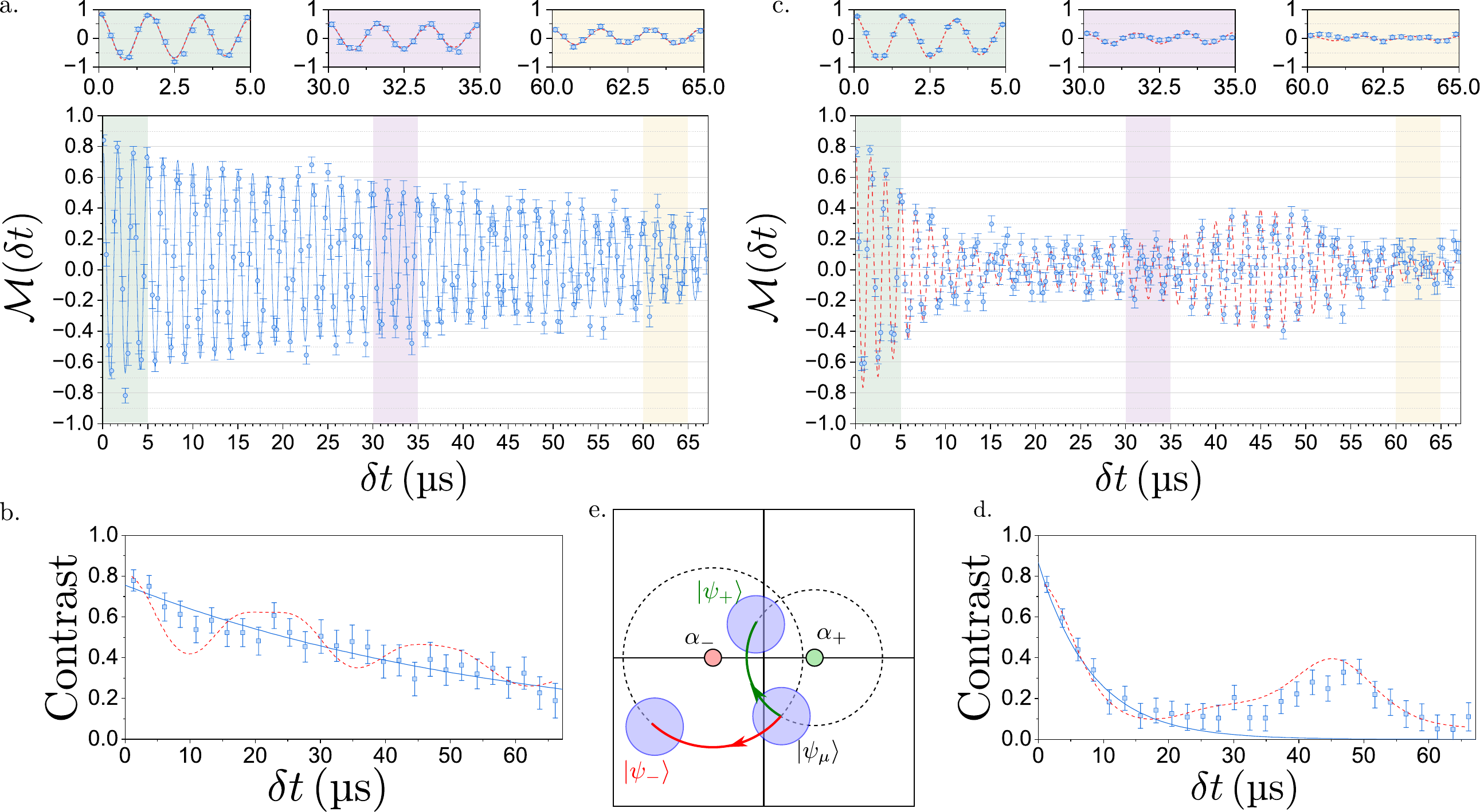}
\caption{\textbf{Long-time spin-exchange interaction and spin-motion coupling.} 
	\textbf{a. Long-duration spin-exchange oscillations.} Population imbalance $\mathcal{M}(\delta t)$ (blue points) measured after the sequence of Fig.~\ref{fig:circuit}.c. The insets are zooms on the colour-shaded time intervals. The solid blue line is an exponentially-damped sinusoidal fit to the data, with a characteristic decay time $T_1 = \SI{59\pm3}{\micro\second}$. The dashed red lines in the insets display the semi-classical numerical simulations of the interaction. 
	\textbf{b.} The contrasts of the measured oscillations (blue squares), of the numerical simulations (red dashed line) and of the exponentially-damped sinusoidal fit to the data of panel a are plotted as a function of $\delta t$. The blue squares and the red dashed line are the results of a fit of a sine wave to the data and numerical simulation of panel a, respectively, over a $\SI{2.1}{\micro\second}$-long moving window. 
	\textbf{c., d. Spin-motion coupling.} Population imbalance (c) and its contrast (d) measured with a $3.3$ times lower trapping power for the circular qubits than for the data of panel a. The colour codes are identical to those of panels a and b. The solid blue line in panel d is the contrast of an exponentially damped sinusoidal fit to the data of panel c with $\delta t \leq \SI{15}{\micro\second}$. The errors bars in panels a and c are Clopper-Pearson confidence intervals, those in panels b and d are standard errors from the fitting procedure. 
	\textbf{d. Phase-space trajectory of the relative motion.} An initial coherent state (lilac-shaded unit-radius disks) $\ket{\psi_\mu}$ rotates in phase space around $\alpha_+$ (green disk) and $\alpha_-$ (red disk) when the spin state of the circular qubits is $\ket{+}$ (green trajectory) and $\ket{-}$ (red trajectory), respectively. After some time, the resulting coherent states $\ket{\psi_\pm}$ differ, leading to a reduction of the contrast of the spin-exchange oscillations by $|\braket{\psi_+|\psi_-}|^2$. For clarity, we chose $|\alpha_\pm|=3.5$ for the plot, while $|\alpha_\pm|=0.08$ or $0.2$ in our experiments, for the high- and low-power BoBs, respectively.}
\label{fig:spin-exchange}
\end{figure*}

The mere operation of the c-CNOT gate would allow us to detect non-destructively atoms in $\kom$ but not to distinguish atoms in $\kim$ from, e.g., state-preparation errors: In both cases, the auxiliary atom is not detected. Because of the $\sim\SI{70}{\percent}$ preparation efficiency of the circular state $\ket{\Sim}$~\cite{Ravon2023}, $\mathrm{C}_1$ and $\mathrm{C}_2$ are prepared in $\ket{\Sos,\Sis}$ in less than $\SI{50}{\percent}$ of the experimental realizations. This would strongly reduce the contrast of spin-exchange oscillations. Thus, we use a syndrome measurement that eliminates preparation errors through post-selection.

\section*{Syndrome measurement}
The syndrome measurement relies on a modification of the auxiliary-atom-based measurement process. Before the fluorescence measurement of the auxiliary qubit state and after the c-CNOT gate, we perform an additional a-CNOT gate. The sequence of c-CNOT and a-CNOT gates is equivalent to an a-CNOT gate followed by a SWAP gate between the circular and auxiliary qubits (inset of Fig.~\ref{fig:circuit}.c). With $\mathrm{A}_k$ initially in $\kia$ and $\mathrm{C}_k$ in $\ket{i_m}, i\in\{0,1\}$, the a-CNOT gate performs a NOT gate on $\mathrm{C}_k$ ($\ket{i_m}\to\ket{(1-i)_m}$). The SWAP gate transfers $\mathrm{A}_k$ to $\ket{(1-i)_a}$ and leaves $\mathrm{C}_k$ in $\kim$ whatever $i$. If $\mathrm{C}_k$ does not initially lie in the measurement basis, it is not affected by the gates and does not end up in $\kim$. %

We detect circular atoms in $\kim$ through a destructive measurement process~\cite{Ravon2023, Machu2026} that maps $\kim$ back to the $\ket{5S}$ ground state. Atoms in $\ket{5S}$ are eventually detected via the same fluorescence measurement than the one used for the auxiliary atoms. The detection of both circular qubits by the destructive measurement ensures a faithful state preparation. Therefore, a post-selection on the measurement of $\mathrm{C}_1$ and $\mathrm{C}_2$ in $\kim$ keeps only the experimental realizations without preparation errors or where the atoms have not jumped out of the measurement or simulation basis. Though the detection efficiency is limited to $\sim\SI{70}{\percent}$, it is similar for $\kom$ and $\kim$ (see Extended Fig.~\ref{edfig:flipflop_noPS}.a). The syndrome measurement does not bias the auxiliary-atom-based measurement.

We plot in Fig.~\ref{fig:circuit}.d the population imbalance $\mathcal{M}(\delta t) = \mathcal{P}_1(\delta t) - \mathcal{P}_2(\delta t)$ obtained after an average over all experimental realizations (orange squares) and over the post-selected ones only (blue circles). Oscillations of $\mathcal{M}(\delta t)$ are conspicuous, revealing the spin-exchange interaction between the circular Rydberg atoms. 
The fitted frequency $J=\SI{594\pm1}{\kilo\hertz}$, in good agreement with the theoretical predictions, corresponds to a real interatomic distance $d=\SI{18.6}{\micro\meter}$. The small discrepancy with the programmed distance stems from trap-array preparation errors~\cite{Mehaignerie2025}.
It is also apparent that the post-selection enables a significant increase of contrast, from $\SI{33\pm1}{\percent}$ to $\SI{79\pm3}{\percent}$. This is consistent with a strong reduction of preparation errors thanks to the syndrome measurement. 
The contrast of the post-selected data is limited by measurement errors which stem from partial loss of the auxiliary atoms from the traps, the finite lifetime of the low-$\ell$ Rydberg level $\koa$, excited during the measurement process, and the efficiency of the optical $\pi$ pulses of the c-CNOT gate. 

\section*{Long-duration spin-exchange dynamics}
We record the same oscillations as in Fig.~\ref{fig:circuit}.d over long times, up to $\delta t = \SI{67}{\micro\second}$ and plot $\mathcal{M}(\delta t)$ in Fig.~\ref{fig:spin-exchange}.a. Spin-exchange oscillations are conspicuous over more than 40 periods. The reduction of contrast with $\delta t$, plotted in Fig.~\ref{fig:spin-exchange}.b, is compatible with an exponential decay (solid line) with a characteristic time $T_1=\SI{59\pm3}{\micro\second}$, an order of magnitude longer than the state of the art~\cite{Emperauger2025a}. 

Half of the measured decay rate stems from the finite lifetime of circular Rydberg levels, $\tau=\SI{145}{\micro\second}$ for $\kis$ at room temperature. Circular Rydberg states $\kcirc{n}$ decay by jumping to $\kcirc{(n\pm1)}$ with almost equal rates. Because half of these quantum jumps, namely $\kos=\kcirc{54}\to\kcirc{55}$ and $\kis=\kcirc{53}\to\kcirc{52}$, make the qubit state leak out of the simulation basis, the syndrome measurement detects and rejects them. 
By cancelling half of the leakage channels, our method effectively lengthens the qubit lifetime. This is confirmed by an analysis of $\mathcal{M}(\delta t)$ measured without post-selection on the syndrome measurement outcome (see Extended Fig.~\ref{edfig:flipflop_noPS}.b). Eventually, relaxation contributes to the reduction of the contrast of the spin-exchange oscillations between the two circular qubits with a characteristic time $T_\mathrm{r} \approx \tau'/2$, where the one-atom lifetime $\tau'$ effectively amounts to $\sim 2\tau$ with the syndrome measurement. With a temperature of $\SI{300}{\kelvin}$, $T_\mathrm{r}=\SI{127}{\micro\second}$ through numerical simulations (see methods). 

\section*{Spin-motion coupling}
The relaxation-induced $1/T_\mathrm{r}$ damping rate of the spin oscillations is too weak to explain the measured $1/T_1$ rate. We attribute the remaining $\approx 1/(\SI{110}{\micro\second})$ contrast-reduction rate to the coupling between the spin and motional degrees of freedom of the interacting circular Rydberg atoms. This coupling originates from the spatial dependency of the interaction strength $J(d)$~\cite{Mehaignerie2023}. The Hamiltonian $\hat{H}_{XY}$~\eqref{eq:HXY} results in a spin-dependent force between the two atoms, repulsive or attractive when the circular qubit pair is in the superpositions $\ket{+}=(\ket{\Sos,\Sis}+\ket{\Sis, \Sos})/\sqrt{2}$ or $\ket{-}=(\ket{\Sos,\Sis}-\ket{\Sis, \Sos})/\sqrt{2}$, respectively. 
In a simplified model that considers the individual trapping potentials to be harmonic with a common frequency $\nu_t$, the spin-motion coupling affects the relative motion only. Linearizing the spatial dependency of $J(d)$, the fictitious particle with reduced-mass $\mu$ oscillates in harmonic potentials of frequency $\nu_t$ but displaced in opposite directions for the spin states $\ket{\pm}$, by $2 x_0 \,\alpha_\pm$, where $x_0=\sqrt{\hbar/(4 \pi \mu \nu_t)}$ is the motional ground-state extension and where $\alpha_\pm = \pm (3x_0/d) \, (J/2\nu_t)$. With an expected oscillation frequency of $\nu_t\approx\SI{34}{\kilo\hertz}$, $|\alpha_\pm|=0.08$. Depending on the spin state $\ket{\pm}$, an initial motional state $\ket{\psi_\mu}$ evolves to $\ket{\psi_\pm}$. The phase-space trajectories are plotted in Fig.~\ref{fig:spin-exchange}.e when $\ket{\psi_\mu}$ is a coherent state and, for clarity, with $|\alpha_\pm|=3.5$.  

With the atoms initially in $\ket{\Sos,\Sis} = (\ket{+}+\ket{-})/\sqrt{2}$, $\hat{H}_{XY}$ prepares the spin-motion entangled state $(\ket{+}\otimes\ket{\psi_+}+\ket{-}\otimes\ket{\psi_-})/\sqrt{2}$. Tracing over the motional state reduces the contrast of the spin-exchange oscillations by $|\braket{\psi_+|\psi_-}|^2<1$, i.e., when the motional states $\ket{\psi_\pm}$ are different enough to provide which-path information about the spin state. At the trap oscillation period, $T_t$, $\ket{\psi_+}=\ket{\psi_-}=\ket{\psi_\mu}$ so that the oscillations are fully contrasted again. This result also holds for the initial thermal state of the atomic motion, which can be represented as a statistical mixture of coherent states.

In order to confirm the influence of motion on the spin exchange, we measure $\mathcal{M}(\delta t)$ after reducing the power of the bottle beams $P_\mathrm{BoB}$ from $P_\mathrm{BoB, H}\approx \SI{0.16}{\watt}$ to $P_\mathrm{BoB, L}=P_\mathrm{BoB, H}/3.3\approx \SI{0.05}{\watt}$. In a harmonic approximation, this reduces the trap frequency to $\nu_t\approx\SI{19}{\kilo\hertz}$ and correspondingly increases the value of $\alpha_\pm$ to $0.2$. The spin-exchange oscillations, plotted in Fig~\ref{fig:spin-exchange}.c, reveal a faster reduction of contrast, plotted in Fig~\ref{fig:spin-exchange}.d, than with a tighter trapping potential, with an initial characteristic decay time $\SI{8.3\pm0.5}{\micro\second}$. Remarkably, a revival of the oscillations is clearly visible at $\delta t\approx\SI{47}{\micro\second}$, close to the expected trap oscillation period of $\SI{53}{\micro\second}$.

We numerically simulate the influence of the atomic motion using a semi-classical model which treats the motion as a classical degree of freedom~\cite{Emperauger2025}. Its predictions exactly match that of a full quantum model for harmonic potentials and Gaussian motional states (see methods). 
We calculate the motion of the atoms during the whole experimental sequence, taking into account the calculated anharmonic three-dimensional shapes of the trapping potentials, from which we obtain the instantaneous value of $J(\delta t)$. We average the resulting signal $\mathcal{M}(\delta t)$ over the initial position and momentum distribution of the atoms in the Gaussian optical tweezers, before their transfer to the BoBs. This distribution is controlled by the temperature $\Theta$ of the atoms, assumed to be the same in all three directions ($\Theta=\SI{9.5}{\micro\kelvin}$ and $\SI{5.9}{\micro\kelvin}$ for the high- and low-power experiments, respectively, see methods). The thermal spread broadens the distribution of interaction strengths, the average over which results in a faster damping of the spin-exchange contrast, while still leading to a revival at the trap oscillation period~\cite{Emperauger2025a}. 
We also take into account the relaxation-induced decay rate $1/T_\mathrm{r}$. 

We fit this model to the data of Fig.~\ref{fig:spin-exchange}.a and c, keeping as free parameters $d$, $P_\mathrm{BoB, H}$ and phenomenological offset and amplitude coefficients to account for experimental imperfections (see methods).
We find a very good agreement between our measurements and the fitted curves (red dashed lines in Fig.~\ref{fig:spin-exchange}) for $d=\SI{18.61}{\micro\meter}$ and $P_\mathrm{BoB, H}=\SI{181}{\milli\watt}$ (hence $P_\mathrm{BoB, L} = P_\mathrm{BoB, H}/3.3 = \SI{54.7}{\milli\watt}$), corresponding to oscillation frequencies at the bottom of trap of $\nu_t=\SI{36}{\kilo\hertz}$ and $\SI{20}{\kilo\hertz}$ with $P_\mathrm{BoB, H}$ and $P_\mathrm{BoB, L}$, respectively. The slight disagreement between the estimated and measured powers may stem from a discrepancy between the calculated and actual profiles of the BoB trapping potential. The model qualitatively reproduces the decay of long-duration spin oscillations with high-power BoBs (Fig.~\ref{fig:spin-exchange}.b). We find, in particular, that the thermal spread along the weakly-trapping laser-propagation direction plays a critical role in the simulated damping. The exaggerated oscillations predicted by the model may indicate a mis-estimation of the temperature for the motion along this direction. However, we find an excellent agreement between the simulated and observed collapse and revival of the oscillations with low-power BoBs (Fig.~\ref{fig:spin-exchange}.c and d). This is remarkable given the few fit parameters and the complexity of the model.

\begin{figure}
\centering
\includegraphics[width=0.95\columnwidth]{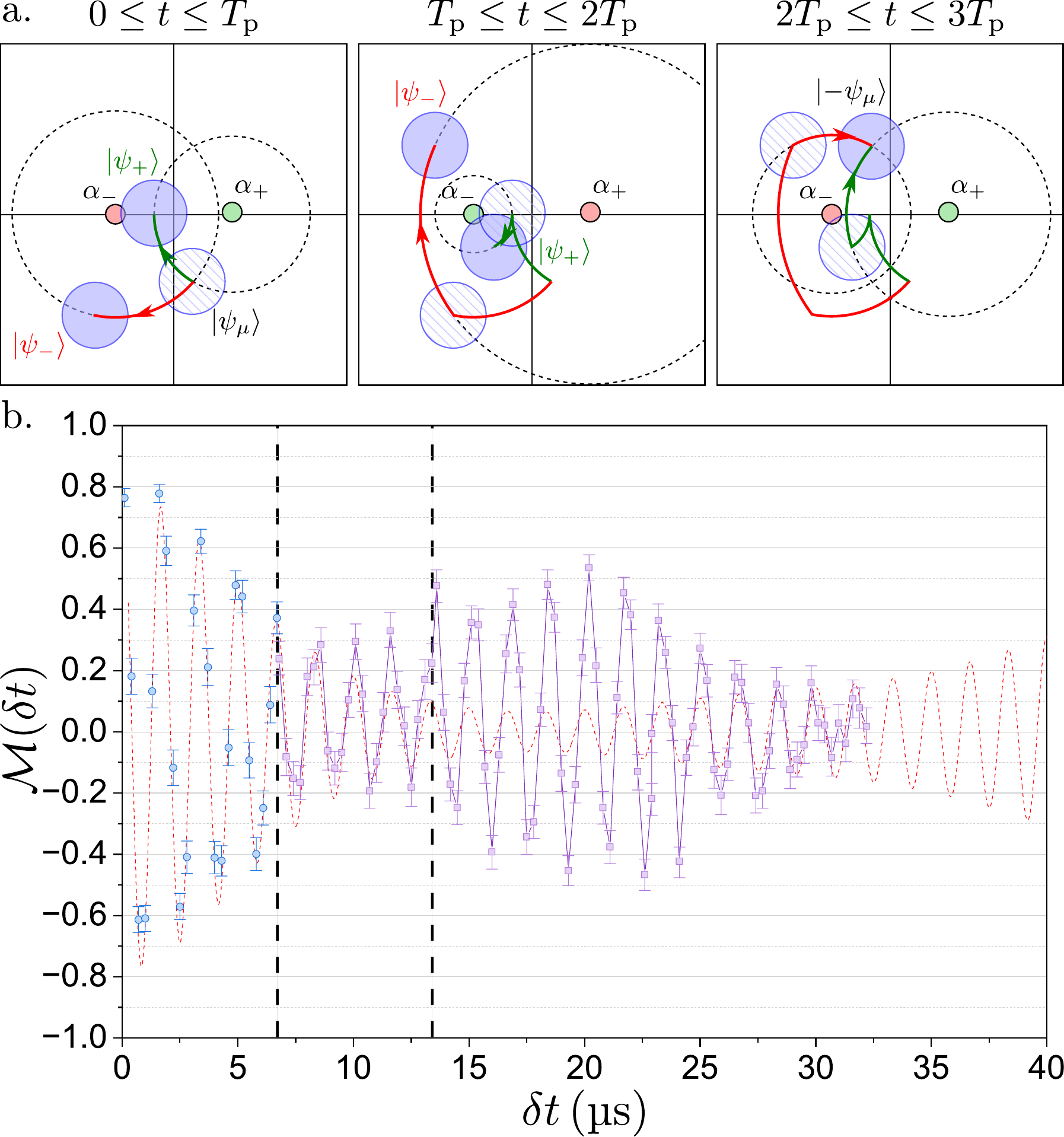}
\caption{\textbf{Dynamical decoupling.} 
		\textbf{a. Phase space trajectories} of an initial coherent state $\ket{\psi_\mu}$ for the relative motion during the dynamical decoupling sequence described in the text. Two phase flips are applied at times $\delta t=T_\mathrm{p}$ and $\delta t=2T_\mathrm{p}$. The trajectories are plotted in red and green for an initial spin state $\ket{+}$ or $\ket{-}$, respectively. The red- and green-shaded disks indicate their respective rotation centres. The three panels (left to right) correspond to the time intervals before the first phase flip ($\delta t < T_\mathrm{p}$, left), between the two phase flips ($T_\mathrm{p} < \delta t < 2 T_\mathrm{p}$, middle) and after the second phase flip ($2 T_\mathrm{p} < \delta t < 3 T_\mathrm{p}$, right) in the case $T_\mathrm{p}=T_t/6$. In all three panels, the dashed and full disks indicate the states at the beginning and end of the time interval, respectively.
		\textbf{b.} \textbf{Population imbalance} $\mathcal{M}(\delta t)$ (connected violet squares) measured with the dynamical decoupling method and $T_\mathrm{p}=\SI{6.7}{\micro\second}$. The times at which we apply the phase flips are indicated by vertical dashed lines. The data for $t<T_\mathrm{p}$ (blue circles) are identical to those of Fig.~\ref{fig:spin-exchange}.c. The red dashed line corresponds to the numerical simulation plotted in Fig.~\ref{fig:spin-exchange}.c. The error bars are Clopper-Pearson confidence intervals.}
\label{fig:echo}
\end{figure}

\section*{Dynamical decoupling of motion}
The measurements of Fig.~\ref{fig:spin-exchange} reveal that long-duration high-contrast spin-exchange oscillations may be obtained by reducing $|\alpha_\pm|$. This can be achieved with higher trap frequencies $\omega_t$ or weaker interaction strength $J$. The former requires increased optical power for the trapping beams while the latter comes at the expense of slower interactions. Alternatively, we devise a method to counteract the effect of motion and the collapse of the spin-exchange oscillation contrast. By letting the auxiliary qubit $\mathrm{A}_2$ perform a $2\pi$ rotation on the $\kia\to\koa$ transition, we apply a $\pi$ phase flip on the circular Rydberg qubit $\mathrm{C}_2$ only~\cite{Machu2026}. This phase rotation transforms $\ket{\pm}$ into $-\ket{\mp}$, turning an initially repulsive (attractive) interactions between $\mathrm{C}_1$ and $\mathrm{C}_2$ into an attractive (repulsive) one. In phase space, the motional state that initially rotated around $\alpha_\pm$ follows, after the phase flip, a circle centred on $\alpha_\mp$ (see Fig.~\ref{fig:echo}.a).

The dynamical decoupling method relies on two phase flips, operated at times $T_\mathrm{p}$ and $2\,T_\mathrm{p}$. In the harmonic approximation, when $T_\mathrm{p} = T_t/6$, the initial coherent state $\ket{\psi_\mu}$ becomes $\ket{-\psi_\mu}$ at $t=3\, T_\mathrm{p}$, whether the spin state is $\ket{+}$ or $\ket{-}$ (see Fig.~\ref{fig:echo}.a), so that $\ket{\psi_+}$ and $\ket{\psi_-}$ overlap again ($|\braket{\psi_+|\psi_-}|^2=1$). This leads to a revival of the spin-exchange oscillations at $t=T_t/2$, half an oscillation period earlier than without the applied spin flips. We perform the decoupling with low-power BoBs, namely with the same experimental parameters as those of Fig.~\ref{fig:spin-exchange}.c. In Fig.~\ref{fig:echo}.b, we plot $\mathcal{M}(\delta t)$ measured with $T_\mathrm{p}=\SI{6.7}{\micro\second}$, for which we expect an overlap between the $+$ and $-$ states of $\SI{99}{\percent}$ at $t=3T_\mathrm{p}$. After the initial reduction of contrast for $t<T_\mathrm{p}$, we observe the expected revival of oscillations with a maximum contrast of $\lesssim \SI{50}{\percent}$ at $t\approx\SI{20}{\micro\second}$. Remarkably, at that time, the contrast of the oscillations was minimal without the phase flips.

\section*{Discussion and outlook}
Our work constitutes the first observation of interaction dynamics between circular Rydberg atoms. It is a decisive step towards quantum simulation with large long-lived arrays of interacting circular Rydberg atoms. While the duration of the measured oscillations is already an order of magnitude beyond the state of the art, the performances could be further improved by operating in a cryogenic setup~\cite{Schymik2021, Zhang2025a, Jin2026}, with circular-Rydberg-level lifetimes in the millisecond range~\cite{Cantat-Moltrecht2020}. To reduce the effect of spin-motion coupling, one could laser cool the atoms to their motional ground state~\cite{Thompson2013}. In a 1D model, the contrast reduction falls below $\SI{3}{\percent}$ with the high-$P_\mathrm{BoB}$ value of $|\alpha_\pm|=0.08$~\cite{Mehaignerie2023}. Quantum simulation with large arrays would require a reduced power per BoB with respect to the $\sim\SI{0.2}{\watt}$ used in this work. Benefiting from the $J/{\nu_t}^{3/2}$ scaling of $\alpha_\pm$, we find that operating with $\SI{20}{\milli\watt}$ per BoB and $J=\SI{25}{\kilo\hertz}$ would already make the contrast reduction due to spin-motion coupling negligible for atoms cooled down to $\SI{1}{\micro\kelvin}$. This would open the route to quantum simulation over milliseconds and the observation of several hundreds of spin-oscillation periods. 

Our results could have broader implications for Rydberg-based quantum simulation and computation. We observe a dramatic increase of simulation time with respect to the state of the art by employing laser-trapped circular Rydberg atoms. The latter, in our room-temperature setup, live as long as the low-$\ell$ level $\ket{70S}$, which can be readily prepared in Rydberg-based quantum computation platforms~\cite{Leclerc2026}. Our work demonstrates that the performance of quantum simulators which employ low-$\ell$ Rydberg levels could be significantly increased by trapping the atoms during the whole simulation. This could be achieved by trapping both the Rydberg and ground-state levels of alkali atoms in repulsive potentials~\cite{Zhang2011} or by using two-electron atoms laser-trapped through their ionic core~\cite{Wilson2019, Holzl2024}. We also show that one can lengthen the effectively lifetime of the Rydberg levels via the syndrome measurement. It suppresses quantum jumps between Rydberg levels, which cannot be corrected by erasure conversion and detection~\cite{Ma2023, Scholl2023}. Thus, it could help reduce error rates in quantum computing platforms.

Finally, the coupling we observe between spin and motion through dipole-dipole interaction between laser-trapped atoms opens the way to the implementation of several proposals, including the  preparation of quantum states of motion~\cite{Mehaignerie2023, Wussler2026} which could be used as a powerful resource~\cite{Shaw2025} or the observation of multi-body interaction~\cite{Nill2025} and molecular dynamics~\cite{Magoni2023}. Interestingly, similar coupling could be observed with laser-trapped molecules, either in full molecular systems~\cite{Anderegg2019, Bao2023, Holland2023, Holman2026} or hybrid Rydberg-molecule platforms~\cite{Zhang2022, Wang2022, Zhu2025, Ruttley2026}.

\section*{Methods}
\subsection*{Experimental system, state preparation and measurement}
We use the experimental platform that has already been described in Refs.~\cite{Ravon2023, Machu2026} and which we sketch in Extended Fig.~\ref{edfig:setup}. We stochastically load Rb-87 atoms in an array of Gaussian optical tweezers (loading array) from a 3D magneto-optical trap (MOT). The loading array of nine optical tweezers ($820$-$\si{\nano\meter}$ wavelength) is generated at the focus of an aspherical lens (Asphericon AFL12-15-P-U-285), with an optical power per trapping site of $\SI{3}{\milli\watt}$. We rearrange the atoms with moveable optical tweezers controlled by an acousto-optic deflector (AOD AA DTS-XY-250) into the array of four working sites sketched in Extended Fig.~\ref{edfig:setup}. Then, we adiabatically reduce the trapping power to cool down the atoms to a measured temperature of $\Theta=\SI{9.5}{\micro\kelvin}$ with a power $P_l=\SI{0.86}{\milli\watt}$ (data of Fig.~\ref{fig:spin-exchange}.a and b) and $\Theta=\SI{5.9}{\micro\kelvin}$ with $P_l=\SI{0.33}{\milli\watt}$ (data of Fig.~\ref{fig:spin-exchange}.c and d and of Fig.~\ref{fig:echo}). The temperatures are measured through a release-recapture experiment fitted to a Monte-Carlo simulation that assumes identical temperatures along the three trap-oscillations directions.

We optically pump the atoms into $\ket{5S_{1/2}, F=2, m_F=2}$ in a $\SI{14}{\gauss}$ magnetic field, aligned with the $x$ axis (axes orientations in Extended Fig.~\ref{edfig:setup}), and kept for all subsequent operations. We excite $\mathrm{C}_1$ and $\mathrm{C}_2$ to the $\ket{52D_{5/2}, m_J=5/2}$ Rydberg level through a $\SI{0.9}{\micro\second}$-long two-photon excitation with global addressing laser beams ($\SI{420}{\nano\meter}$ and $\SI{1015}{\nano\meter}$ wavelengths). To do so, we effectively turn off the optical tweezers for these two atoms only by transferring the auxiliary atoms to the array of auxiliary sites (power $P_a=\SI{1}{\milli\watt}$ per tweezer) which we turn on $\SI{10}{\micro\second}$ before switching off the loading array (see Extended Fig.~\ref{edfig:sequence} for a sketch of the experimental sequence). We perform the Rydberg excitation $\SI{0.6}{\micro\second}$ later. The trap-induced light shifts of $\SI{8}{\mega\hertz}$ prevent the excitation of $\mathrm{A}_1$ and $\mathrm{A}_2$. We turn on the array of optical bottle beams to trap both Rydberg atoms $\SI{0.4}{\micro\second}$ after the Rydberg excitation, and transfer the atoms to the circular Rydberg level $\kim=\kcirc{52}$ via microwave and radiofrequency fields~\cite{Ravon2023}. The state preparation, from the optical excitation of $\ket{52D_{5/2}, m_J=5/2}$ to the preparation of $\kim$, lasts $\SI{9}{\micro\second}$ and has an efficiency of $\sim\SI{70}{\percent}$. The initially vanishing electric field is turned on in the magic configuration (along the $x$ axis) during the circular state preparation. Its final value is $\SI{1.59}{\volt\per\centi\meter}$.

The power of the auxiliary trapping beams is reduced from $P_a$ to $P_l$ immediately after the optical excitation of $\ket{52D_{5/2}, m_J=5/2}$. During the sequence, the auxiliary atoms are transiently excited to $\koa=\ket{45S_{1/2}, m_J=1/2}$ through a $\SI{0.3}{\micro\second}$-long two-photon excitation with global addressing laser beams ($\SI{420}{\nano\meter}$ and $\SI{1017}{\nano\meter}$ wavelengths). The auxiliary array is turned off during the excitation of both atoms. Selective excitation of the auxiliary atom $\mathrm{A}_2$ is achieved by shifting within $~\SI{3}{\micro\second}$ the auxiliary array with the AOD, as indicated in Extended Fig.~\ref{edfig:setup} and described in Ref.~\cite{Machu2026}, while simultaneously increasing its power back to $\SI{1}{\milli\watt}$. Auxiliary atom $\mathrm{A}_1$ only is trapped during the laser excitation and shielded from the laser excitation by the trap-induced light shifts.

All atoms in the $5S$ ground state ($F=1$ and $F=2$ manifolds) are detected by fluorescence measurements on the $\ket{5S_{1/2}, F=2, m_F=2}\to\ket{5S_{1/2}, F'=3, m_F=3}$ transition with the 3D MOT laser beams and a repumping beam, resonant on the $F=1\to F'=2$ transition, shone during $\SI{20}{\milli\second}$. During the fluorescence measurement, the power of the Gaussian optical tweezers is increased to $\SI{3}{\milli\watt}$.

\subsection*{c- and a-CNOT gates}
The Förster resonance between the circular-auxiliary atom pair states $\ket{54\mathrm{C}}\otimes\ket{45S_{1/2}, m_J=1/2}$ and $\ket{53\mathrm{C}}\otimes\ket{45P_{3/2}, m_J=3/2}$ in a $F=\SI{1.59}{\volt\per\centi\meter}$ electric field results in a first-order dipole-dipole interaction between two atoms in $\koa$ and $\kom$~\cite{Machu2026}, on which rely the c-CNOT and a-CNOT gates. 

For the c-CNOT gate, we transiently optically excite the auxiliary atoms from $\kia$ to the low-$\ell$ Rydberg level $\koa$. Rydberg blockade prevents this excitation in the presence of a circular Rydberg atom in $\kom$. Auxiliary atoms left in $\kia$ are optically pumped to $\ket{5S_{1/2}, F=1}$ while the excited ones are de-excited to $\kia$ and expelled from the trap with a push-out laser beam at $\SI{780}{\nano\meter}$, resonant on the $\kia\to\ket{5P_{3/2}, F'=3, m_F=3}$ transition. The final fluorescence measurement detects all atoms in their ground state, effectively measuring the population in $\kia$ after the first optical excitation. It amounts to a non-destructive measurement of the population in $\kom$ for the circular Rydberg atom. Destructive measurement of the population in $\kim$ is performed by de-exciting the circular Rydberg atom back to $\ket{5S_{1/2}, F=2, m_F=2}$ before the final fluorescence measurement~\cite{Ravon2023}.

Similarly to the c-CNOT gate, the a-CNOT gate relies on Rydberg blockade. We perform a $\SI{0.8}{\micro\second}$-long mw $\pi$ pulse on the $\kim\to\kom$ transition. The mw transition is blocked when the neighbouring auxiliary atom has been excited to $\koa$, via the X gates performed during  selective excitation or during the c-CNOT gate applied before the a-CNOT gate (see inset of Fig.~\ref{fig:circuit}.c). We apply the mw pulse $\SI{0.4}{\micro\second}$ after the optical pulse on the auxiliary atoms (see Extended Fig.~\ref{edfig:sequence} for the detailed timings).

\subsection*{Interaction strengths}
The strength of the interaction between two circular Rydberg atoms in $\kis$ and $\kos$ takes the simple form of Eq.~\eqref{eq:HXY}, with $J(d)=\SI{3847}{\giga\hertz\cdot\micro\meter^3}/d^3$ as calculated from a hydrogenic formula~\cite{Mehaignerie2025}. When we rotate the electric field and modify the angle $\theta$ it forms with the interatomic axis, the interaction strength scales as $3\cos^2\theta-1$ even while keeping the magnetic field along the $\theta=0$ direction. This stems from the fact that the electric field essentially sets the quantization axis for circular Rydberg states, given the electric field of $\sim\SI{1}{\volt\per\centi\meter}$ and the magnetic field of $\SI{14}{\gauss}$~\cite{Mehaignerie2023}.

The situation is more complex for the interaction between the circular and auxiliary atoms, the quantization axis of the latter being controlled by the magnetic field. This interaction strength, denoted $J_\mathrm{AC}$, is maximal when $\theta=\theta_\mathrm{M}$ (interatomic axis parallel to the electric and magnetic field) but its dependency with $\theta$ is more complex than Eq.~\eqref{eq:HXY}. With an interatomic distance of $a=\SI{5.2}{\micro\meter}$, when $\theta=\theta_\mathrm{M}$, we expect~\cite{Mogerle2026} $J_\mathrm{AC}=\SI{32}{\mega\hertz}$ between $\mathrm{A}_1$ and $\mathrm{C}_1$ and a negligible sub-kilohertz interaction between $\mathrm{A}_1$ and $\mathrm{C}_2$. Experimentally, from the splitting of the $\kos$ optical excitation spectrum~\cite{Machu2026}, we measure $\SI{34}{\mega\hertz}$ and $\SI{28}{\mega\hertz}$ for the $\mathrm{A}_1\mathrm{C}_1$ and $\mathrm{A}_2\mathrm{C}_2$ pairs, respectively, indicating an actual value of $a$ of $\SI{5.1}{\micro\meter}$ and $\SI{5.4}{\micro\meter}$.

\renewcommand{\tablename}{Extended Tab.}
\begin{table}
\centering
\begin{tabular}{c|c||c|c}
$(i,j,k)$ & ${a_{ijk}}/{\mathrm{k}_\mathrm{B}} $ & $(i,j,k)$ & ${a_{ijk}}/{\mathrm{k}_\mathrm{B}}$ \\
\hline
$(0,0,1)$ & $0.292$ & $(1,0,0)$ & $1.50$ \\
$(0,0,2)$ & $-1.69\cdot10^{-3}$ & $(1,0,1)$ & $-0.371$\\
$(0,0,3)$ & $-1.74\cdot10^{-6}$ & $(1,1,0)$ & $3.375$\\
$(0,1,0)$ & $1.36$  & $(1,0,2)$ & $1.53\cdot10^{-3}$\\
$(0,1,1)$ & $-0.375$ & $(1,1,1)$ & $0.190$ \\
$(0,1,2)$ & $1.56\cdot10^{-3}$ & $(1,2,0)$ & $-2.34$ \\
$(0,2,0)$ & $2.07$ & $(2,0,0)$ & $1.82$ \\
$(0,2,1)$ & $0.101$ & $(2,0,1)$ & $9.97\cdot10^{-2}$\\
$(0,3,0)$ & $-0.974$ & $(2,1,0)$ & $-2.26$ \\
		  &			 & $(3,0,0)$ & $-0.887$
\end{tabular}
\caption{Table of $a_{ijk}$ coefficients used in the polynomial approximation of the numerically calculated trapping potential for $\kim$. The values of $a_{ijk}/\mathrm{k}_\mathrm{B}$, where $\mathrm{k}_\mathrm{B}$ is Boltzmann's constant, are given in $\si{\micro\kelvin\per\micro\meter^{2(i+j+k)}\per\milli\watt}$. }
\label{edtab:interactions}
\end{table}

The van der Waals spin-exchange interaction between atoms $\mathrm{C}_1$ and $\mathrm{C}_2$ when $\theta=\theta_\mathrm{M}$ amounts to $\SI{0.17}{\kilo\hertz}$~\cite{Nguyen2018}. The van der Waals shift of the $\kia\to\koa$ optical transition in the vicinity of a circular Rydberg atom in $\kim$ equals $\SI{156}{\kilo\hertz}$.

\subsection*{Phase flips and dynamical decoupling}
By performing a $2\pi$ pulse on the $\kia\to\koa$ transition on $\mathrm{A}_2$, initially in $\kia$, and trapped in the vicinity of $\mathrm{C}_2$ in $\ket{i_m}$, we apply a $\pi$ phase shift on the $\mathrm{A}_2$-$\mathrm{C}_2$ pair state $\kia\otimes\ket{i_m}$ unless $i=0$. In the latter case, Rydberg blockade prevents the $2\pi$ rotation and so the phase shift. Because $\mathrm{A}_2$ always ends up in $\kia$, this effectively performs a phase flip on the circular Rydberg qubit $\mathrm{C}_2$~\cite{Machu2026}. Applying this phase shift on $\mathrm{C}_2$ only transforms $\ket{\pm}=(\ket{\Sos,\Sis}\pm\ket{\Sis, \Sos})/\sqrt{2}$ into $-\ket{\mp}=(-\ket{\Sos,\Sis}\pm\ket{\Sis, \Sos})/\sqrt{2}$. During the dynamical decoupling sequence, to apply such a phase flip, we first rotate the electric field back to $\theta=\theta_\mathrm{M}$ to freeze the spin-exchange interaction, perform the $2\pi$ optical excitation of $\mathrm{A}_2$, whose trap is off, and eventually turn the electric field back to $\theta=\pi/2$ to restore the spin-exchange interaction. The effective total duration of the phase flip is $\tau_\mathrm{p} = \SI{3}{\micro\second}$. The electric-field rotation, made within $~\SI{0.1}{\micro\second}$, is fast enough with respect to the spin-exchange interaction frequency of $\SI{0.6}{\mega\hertz}$ to neglect interactions between the $\mathrm{C}_1$ and $\mathrm{C}_2$, so that we do not need to take into account the $\tau_\mathrm{p}$ delay in the calculation of the effective interaction time, $\delta t$.  

\subsection*{Numerical simulations}
The numerical simulations of Fig.~\ref{fig:spin-exchange} are Monte-Carlo simulations of the trajectories of the circular Rydberg atoms in their traps, similar to those performed in Ref.~\cite{Emperauger2025a}. We randomly initialize the position and speed of an atom according to its thermal spread in the initial Gaussian optical tweezers approximated by three-dimensional (3D) harmonic potential, taking into account the measured values of the temperature $\Theta$ and trapping power $P$. We calculate the 3D trajectory of an atom, taking into account the numerically-estimated 3D shape of the bottle beam traps. The latter is calculated by averaging the ponderomotive potential deduced from the BoB intensity profile over the $\kim$ wavefunction. It is approximated by a six-order 3D polynomial 
\begin{align}
	U(x,y,z ; t) = P_\mathrm{BoB}(t) \, \sum_{1\leq i+j+k\leq3} a_{ijk} \, \tilde{x}^{2i} \tilde{y}^{2j} z^{2k} \ ,
\end{align}
where $P_\mathrm{BoB}(t)$ is the BoB instantaneous power, the $z$ axis in the BoB propagation direction (see Extended Fig.~\ref{edfig:setup}), the $\tilde{x}$ axis is the $\mathrm{C}_1\mathrm{C}_2$ axis and the $\tilde{y}$ axis is the direction of the electric field when $\theta=\SI{90}{\degree}$. The $a_{ijk}$ coefficients are given in Extended Table~\ref{edtab:interactions}. 

In the calculation of the trajectories, we consider the time intervals during which the potential is turned off and we disregard the dipole-interaction force~\cite{Emperauger2025a}. With two different trajectories, we compute the relative position of two atoms from which we calculate the instantaneous strength $J(\delta t)$ of their dipole-dipole interaction, taking into account both the interatomic distance and the orientation of the interatomic axis with respect to the electric field. We compute $190$ single-atom trajectories from which we get $N_t = 190^2=36100$ relative trajectories and $N_t$ realizations of $J(\delta t)$. The output of the function used to fit the data in Fig.~\ref{fig:spin-exchange}.a and c is the average 
\begin{align}
	\mathcal{M}(\delta t) = \mathcal{M}_0 + A \, \langle\cos\left[2\pi \, J(\delta t) \delta t\right]\rangle_{N_t} \times \e^{-t/T_\mathrm{r}} \ , \label{eq:fitfunction}
\end{align} 
where $A$ and $\mathcal{M}_0$ are the amplitude and offset coefficients which act as free fit parameters (see main text), $\langle\cdot\rangle_{N_t}$ denotes the average over the $N_t$ simulated realizations, and where $T_\mathrm{r}$ is the expected relaxation-induced two-atom characteristic decay time.

To estimate the latter, we numerically simulate with the QuTip package~\cite{Lambert2026} the spin exchange between two circular Rydberg atoms in the presence of relaxation. We perform a Monte-Carlo simulation in the eigenspace of basis $\{\kcirc{n}, 50\leq n\leq57\}$. The jump operators take into account the $\SI{300}{\kelvin}$ mw blackbody radiation and possible collective effects in the relaxation, the atoms $\mathrm{C}_1$ and $\mathrm{C}_2$ being much closer than the involved millimetre-range wavelengths. The atoms are considered fixed and $J$ is set to $\SI{600}{\kilo\hertz}$. Initializing the atoms in $\ket{\Sis,\Sos}$ we calculate the probabilities $\mathcal{P}_{10}$ and $\mathcal{P}_{01}$ to find them in $\ket{\Sis, \Sos}$ and $\ket{\Sos, \Sis}$, respectively, by keeping only the trajectories where both atoms are measured in the simulation basis to reproduce the syndrome measurement. We calculate $\widetilde{\mathcal{M}}(\delta t) = \mathcal{P}_{10}(\delta t) - \mathcal{P}_{01}(\delta t)$ and plot it in Extended Fig.~\ref{edfig:flipflop_noPS}.c. The simulated data are well fitted by an exponentially-damped sinusoidal function, with a characteristic decay time of $\SI{126.8\pm0.1}{\micro\second}$. The two-atom relaxation time, lengthened via the syndrome measurement, is close to the one-atom lifetimes of $\kis$ and $\kos$, of $\SI{145}{\micro\second}$ and $\SI{150}{\micro\second}$, respectively.

We fit Eq.~\eqref{eq:fitfunction} to the data of figures~\ref{fig:spin-exchange}.a and c together with up to $20000$ estimations of a Simplex algorithm. The free parameters are the distance between the centres of the trapping potentials, the power $P_\mathrm{BoB, H}$ used for the data of Fig.~\ref{fig:spin-exchange}.a -- its value for the data of Fig.\ref{fig:spin-exchange}.c being constrained to $P_\mathrm{BoB, L} = P_\mathrm{BoB, H}/3.3$ from the known reduction factor applied experimentally --, and the amplitude and offset coefficients $A$ and $\mathcal{M}_0$. The fitted curves plotted in Fig.~\ref{fig:spin-exchange} are obtained with $d=\SI{18.61}{\micro\meter}$, $P_\mathrm{BoB, H}=\SI{181}{\milli\watt}$, $A=0.82$ and $\mathcal{M}_0=0.05$ for the data of panel a and $P_\mathrm{BoB, L} = \SI{54.7}{\milli\watt}$, $A=0.78$ and $\mathcal{M}_0=0$ for the data of panel c.

\subsection*{Measurements with no syndrome measurement}
We plot in Extended Fig.~\ref{edfig:flipflop_noPS}.b the results of the spin-oscillation measurements performed with high-power BoBs but without the syndrome measurement used to produce the data of Fig.~\ref{fig:spin-exchange}: The population imbalance $\mathcal{M}$ is averaged over all experimental realizations, with no post-selection on the final destructive detection of the circular Rydberg atoms, even if the syndrome measurement sequence is actually performed. The oscillations of $\mathcal{M}(\delta t)$ are still visible but with a reduced contrast, as for the data of Fig.~\ref{fig:circuit}.d. The contrast reduction is well fitted by an exponentially-damped sine with a characteristic decay time of $\SI{36\pm2}{\micro\second}$. We also plot in Extended Fig.~\ref{edfig:flipflop_noPS}.a the probabilities $\mathcal{P}_c$ to recapture $\mathrm{C}_1$ (violet triangles) and $\mathrm{C}_2$ (green triangles) at the end of the sequence. Atom $\mathrm{C}_1$ is better recaptured than $\mathrm{C}_2$, which may stem from residual trap power inhomogeneities or misalignement between the loading and BoB arrays. There are no visible oscillations of the probabilities $\mathcal{P}_c$, indicating that $\kim$ and $\kom$ have nearly equal detection efficiencies. The long-term reduction of $\mathcal{P}_c$ is in qualitative agreement with the decay of the circular Rydberg states (black line). Short-term fluctuations of $\mathcal{P}_c$, which prevent a quantitative fit to an exponential damping, are due to shot-to-shot fluctuations of the experiment.

\subsection*{Equivalence between classical and quantum models}
We treat the motion as a classical degree of freedom in the numerical simulation. This is supported by the fact that the classical~\cite{Emperauger2025a} and quantum~\cite{Mehaignerie2023} models exactly match under the following approximations: we consider a 1D harmonic potential of frequency $\nu_t = \omega_t/2\pi$, we consider a small enough motional extension to linearize the interaction potential, and we assume that the quantum state is a Gaussian state. We provide here elements of demonstration. Considering the initial quantum state $\ket{\psi_\mu}$ for the relative motion, the quantum state of spin and motion after a time $t$ reads
\begin{align}
\ket{\Psi(t)} &= \frac{\ket{\psi_+(t)}\e^{-i \pi J t}\ket{+} + \ket{\psi_-(t)}\e^{i \pi J t}\ket{-}}{\sqrt{2}} , 
\end{align}
where $\ket{\psi_\pm(t)} = \hU_\pm(t)\ket{\psi_\mu}$, with $\hU_\pm(t)$ the time-evolution operator in a harmonic potential centred on $\alpha_\pm$. Tracing over the motional state, the density matrix of the spin state reads
\begin{align}
\hat{\rho}_\mathrm{sp} &= \frac{1}{2} \left[ \ket{+}\!\bra{+}  \, + \,  \ket{-}\!\bra{-} \vphantom{\hU_-^\dagger} \right.\nonumber \\
						&  + \quad  \e^{-2i\pi Jt}\,\braket{\psi_-(t)|\psi_+(t)} \, \ket{+}\!\bra{-} \nonumber \\
						& \left.+ \quad \e^{+2i\pi Jt}\,\braket{\psi_+(t)|\psi_-(t)} \, \ket{-}\!\bra{+} \right] \ .
\end{align}
In the following, we denote $\mathcal{C}(\ket{\psi_\mu}, t) = \braket{\psi_-(t)|\psi_+(t)} = \bra{\psi_\mu}\hU_-^\dagger(t)\,\hU_+(t)\ket{\psi_\mu}$, $|\mathcal{C}(\ket{\psi_\mu}, t)|$ being the contrast of the recorded spin oscillations with the initial motional state $\ket{\psi_\mu}$.

We now assume that $\ket{\psi_\mu}$ is a coherent state $\ket{\beta}$ where $\beta = \beta_r + i \beta_i$ is a complex number. Its time evolution takes the simple form $\ket{\psi_\pm}(t) = \e^{-i \omega_t t/2 -i \alpha_\pm(B_{\pm,i}-\beta_i)} \ket{B_\pm(t)}$ with $B_\pm(t) = \alpha_\pm + (\beta-\alpha_\pm)\,\e^{-i\omega t}$. In this case
\begin{align}
	\mathcal{C}(\ket{\beta}, t) &= \e^{-4\alpha_+^2 \left(1-\cos \omega_t t\right) + i\Phi(\beta, t)} \ , \label{eq:overlap}
\end{align}
where $\Phi(\beta, t) = 4 \alpha_+ \left[\beta_i(1-\cos\omega_t t) + \beta_r \sin\omega_t t\right]$. Equation~\eqref{eq:overlap} can be recast into the form of an average over the spatial and momentum coordinates $\hat{x}$ and $\hat{p}$ of the fictitious particle via the normalized variables $\hat{X} = \hat{x}/x_0 = \hat{a} + \hat{a}^\dagger$ and $\hat{P} = \hat{p} / \left[\hbar/(2x_0)\right] = i (\hat{a}-\hat{a}^\dagger)$ of average values $2\beta_r$ and $2\beta_i$, respectively, and unit variance for the coherent state $\ket{\beta}$:
\begin{align}
	\mathcal{C}(\ket{\beta}, t) &= \nonumber \\
	&\hspace{-7ex} \iint \frac{\mathrm{d}X\mathrm{d}P}{2\pi} \e^{-[{(X-2\beta_r)^2}+{(P-2\beta_i)^2}]/{2}} \e^{i{\Phi(X+iP,\, t)}/{2}} \ . \label{eq:integral}
\end{align}
Eventually, one finds
\begin{align}
	\mathcal{M}(t) &= \iint \frac{\mathrm{d}X\mathrm{d}P}{2\pi} \e^{-[{(X-2\beta_r)^2}+{(P-2\beta_i)^2}]/{2}} \cos \xi(t) \ , \label{eq:average} \\
	\frac{\xi(t)}{2\pi} &= Jt - \frac{3 J x_0}{d \omega_t}\left[X \sin\omega_t t + P (1-\cos\omega_t t)\right] \nonumber \\
	 &= \int_0^t \mathrm{d}t' \, J \, \left[1-3\frac{x(t')}{d}\right] \ , \label{eq:phase}
\end{align}
where $x(t) = x_0 \left[X\cos\omega_t t + P\sin\omega_t t\right]$ is the instantaneous position of the fictitious particle of initial position $x_0 X$ and velocity $x_0 \omega_t P= [\hbar/(2x_0)] P/\mu$. The argument of the integral in equation~\eqref{eq:phase} is the instantaneous value of the spin-oscillation frequency linearized around the interatomic distance $d$. Thus, the value of $\mathcal{M}(t)$ in Eq.~\eqref{eq:average} is that one would find considering a classical motion of the fictitious particle~\cite{Emperauger2025a}. This results holds whatever the initial coherent state, including the motional ground state, so that the equivalence between classical and quantum models is not limited to highly excited motional states. It is valid for any statistical superposition of coherent states, in particular for thermal states as is the case in our experiments. Averaging Eq.~\eqref{eq:average} over the thermal distribution leads to the results already found in Ref.~\cite{Mehaignerie2023}. Observing a deviation from the classical predictions would require the preparation of a non-classical motional state~\cite{Mehaignerie2023,Wussler2026}. 

\bibliography{Duran2026} 

\newpage

\renewcommand{\figurename}{Extended Data Fig.}
\setcounter{figure}{0}
\renewcommand{\thefigure}{\arabic{figure}}

\begin{figure*}
\centering
\includegraphics[width=0.75\linewidth]{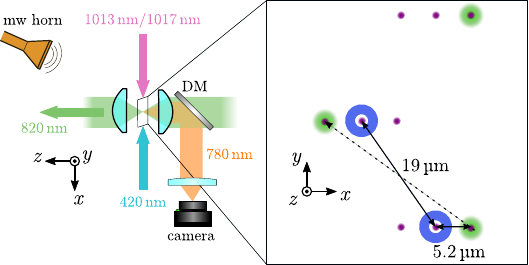}
\caption{\textbf{Sketch of the experimental setup.} Two aspherical lenses (cyan) are used to focus the trapping laser beam (green, $\SI{820}{\nano\meter}$) onto the atoms, and to collect the light out of the experimental chamber. The focusing lens collects fluorescence photons (orange, $\SI{780}{\nano\meter}$) detected by a camera after a dichroic mirror (DM). The global Rydberg excitation lasers (blue $\SI{420}{\nano\meter}$ and pink $1015$ and $\SI{1017}{\nano\meter}$) propagate along the $x$ axis. A horn antenna shines mw radiation in the whole experimental chamber. \textbf{Inset.} We use three different arrays of trapping sites, prepared and overlapped using spatial light modulators: a loading array (violet, nine sites), an auxiliary array (green, three sites) and an array of BoBs (blue, two sites). The dashed arrow indicates the displacement we apply on the auxiliary array for selective excitation of the $\mathrm{A}_2$.}
\label{edfig:setup}
\end{figure*}

\begin{figure*}
\centering
\includegraphics[width=0.75\linewidth]{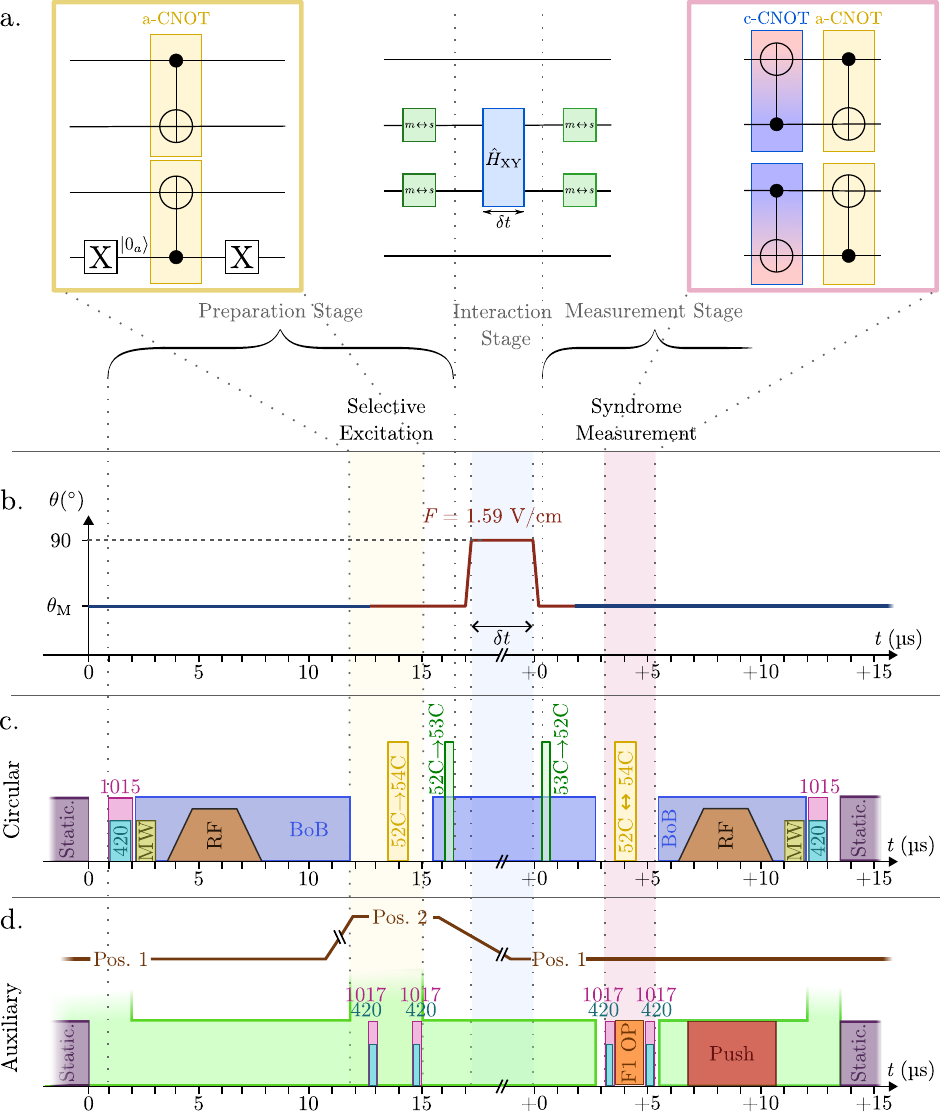}
\caption{\textbf{Sketch of an experimental sequence.} Timings of the experimental sequence used to record the data in Fig.~\ref{fig:spin-exchange}, from the state preparation to the measurement stage. 
	\textbf{Panel a} reproduces parts of the logical sequence sketched in Fig.~\ref{fig:circuit}.c. 
	\textbf{Panel b} represents the value of $\theta(t)$,  modified from $\theta_m$ to $\SI{90}{\degree}$ during the interaction stage (blue box in Fig.~\ref{fig:circuit}). The time during which the value of the electric field is constant equal to $\SI{1.59}{\volt\per\centi\meter}$ is indicated by a red line. The interaction stage, during which $\theta=\SI{90}{\degree}$, lasts $\delta t$. All events that follow the interaction stage have a fixed timing with respect to the end of the latter, they are displaced accordingly when $\delta t$ is modified. The labels with a "$+$" indicate the time delay from the end of the interaction stage. 
	\textbf{Panel c} represents the sequence of optical and microwave fields felt by $\mathrm{C}_1$ and $\mathrm{C}_2$. The violet- and blue-shaded boxes indicate when the loading and BoB arrays are on, respectively. The green- and yellow-shaded boxes represent the mw $\pi$ pulses on the $\kcirc{52}\leftrightarrow\kcirc{54}$ ($\kim\leftrightarrow\kom$) and $\kcirc{52}\leftrightarrow\kcirc{53}$  ($\kim\leftrightarrow\kis$) transitions, respectively. The blue- and pink-shaded boxes represent the blue and infra-red Rydberg excitation lasers used to perform $\pi$-pulses on the $\ket{5S_{1/2}, F=2, m_F=2}\leftrightarrow\ket{52D_{5/2}, m_J=5/2}$ transition. The yellow MW box corresponds to a mw $\pi$ pulse to the intermediate state $\ket{52F_{7/2}, m_J=5/2}$ and the brown trapeze the adiabatic radiofrequency-field-induced transfer to $\kcirc{52}$. 
	\textbf{Panel d} represents the sequence of optical fields felt by the auxiliary atoms. The light-green-shaded boxes represent the time during which the auxiliary array is on. Their heights schematically represent the power per trapping site: It is kept high equal to $\SI{1}{\milli\watt}$ during the laser excitation of $\ket{52D_{5/2}, m_J=5/2}$ (panel b) and during the site-selective excitation, and reduced to $P_l$ otherwise. The blue- and pink-shaded boxes represent the blue and infra-red Rydberg excitation lasers used to perform $\pi$-pulses on the $\ket{5S_{1/2}, F=2, m_F=2}\leftrightarrow\ket{45S_{1/2}, m_J=1/2}$ transition. The orange-shaded box indicates the optical pumping stage to the $F=1$ ground state and the dark-red-shaded box the push-out stage during which remaining atoms in the $F=2$ ground-state manifold are expelled from the traps. The top brown line sketches the position of the auxiliary array, from position 1 for which both auxiliary atoms are trapped to position 2 for which $\mathrm{A}_1$ only is trapped (see also Extended Fig.~\ref{edfig:setup}).}
\label{edfig:sequence}
\end{figure*}

\begin{figure*}
\centering
\includegraphics[width=0.75\linewidth]{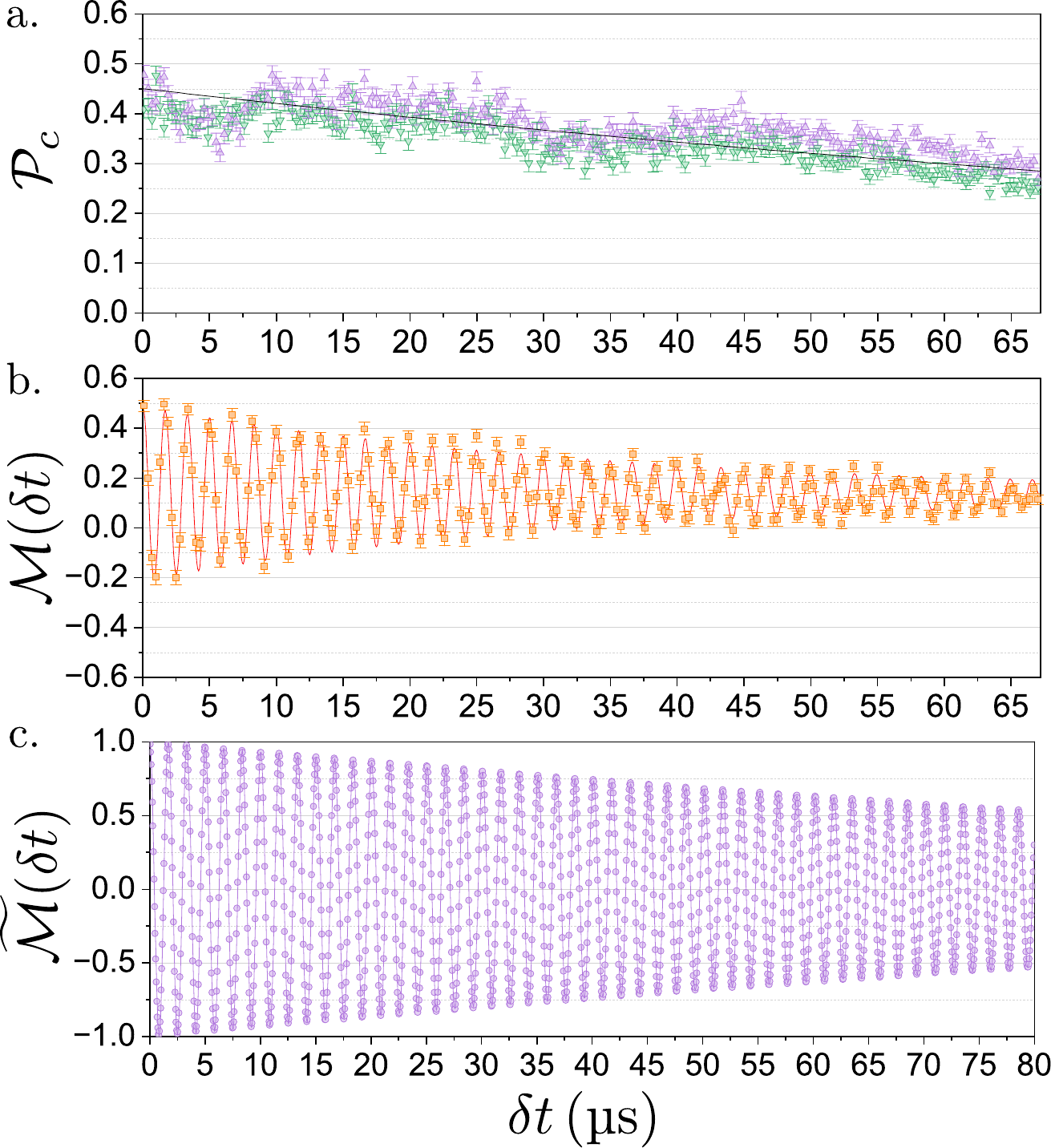}
\caption{\textbf{Measurement without the syndrome measurement.} \textbf{a. Probability to measure the circular Rydberg atoms} in $\kim$ at the end of the sequence sketched in Fig.~\ref{fig:circuit}.c, for atoms $\mathrm{C}_1$ (upper violet triangle) and $\mathrm{C}_2$ (lower green triangle) atoms. The solid black line is an exponential decay with a characteristic decay time of $\SI{147}{\micro\second}$, equal to the average lifetime of $\kcirc{53}$ and $\kcirc{54}$ at $\SI{300}{\kelvin}$. Besides shot-to-shot fluctuation, there is no visible oscillation of probabilities, indicating an equal detection efficiency for $\kim$ and $\kom$. \textbf{b. Population imbalance} $\mathcal{M}(\delta t)$ (orange squares) measured after the sequence of Fig.\ref{fig:circuit}.c with a trapping power $P=P_\mathrm{BoB, H}\approx\SI{0.16}{\watt}$ but without post-selection on the destructive detection of the circular Rydberg atoms. This corresponds to the same data as these plotted as orange squares in Fig.~\ref{fig:setup}.e. The solid orange line is an exponentially-damped sinusoidal fit to the data with a characteristic decay time $T_1=\SI{36\pm1}{\micro\second}$. The error bars are Clopper-Pearson confidence intervals. \textbf{c. Simulated spin-oscillations} in the absence of spin-motion coupling but considering the finite lifetime of circular Rydberg states at room temperature. The simulated data (violet circles) is well fitted by an exponentially-damped sine with a characteristic decay time of $\SI{126.8}{\micro\second}$ and an oscillation frequency of $\SI{598}{\kilo\hertz}$.}
\label{edfig:flipflop_noPS}
\end{figure*}

\subsection*{Acknowledgements}
We acknowledge decisive technical support from Jêrôme Beugnon and PASQAL for lending us lasers. We also acknowledge support from Russell Thomas and Ilian Despard for their help in maintaining our Ti:Saph lasers.

This publication has received funding by the France 2030 programs of the French National Research Agency (Grant No. ANR-22-PETQ-0004, project QuBitAF), under Horizon Europe programme HORIZON-CL4-2022-QUANTUM-02-SGA via the project 101113690 (PASQuanS2.1), by the European Union (ERC Advanced Grant No. 786919, project TRENSCRYBE). It has been supported by the Île-de-France region in the framework of DIM QuanTiP (project LT-CRAQS) and by the Quantum Information Center Sorbonne as part of the program \emph{Investissements d'excellence} -- IDEX of the Alliance Sorbonne Université. 

\subsection*{Author contributions}
A.D.H., G.C., A.A.Y. and A.K. contributed to the experimental setup and performed the experiments. G.C. and C.S. analysed the data. A.A.Y. and A.K. performed the numerical simulations. Y.M. contributed to the experimental setup and to the syndrome measurement protocol. M.B., J.M.R. and C.S. conceived the experiment and supervised the project. All authors contributed to the manuscript. 

\end{document}